# ENHANCING INFRASTRUCTURE SECURITY IN REAL ESTATE

KYLE DEES, M.S.[1] AND SYED (SHAWON) RAHMAN, PH.D.[2]

[1]Information Assurance and Security, Capella University, Minneapolis, MN, USA
Dees.Kyle@gmail.com

[2]Assistant Professor, University of Hawaii-Hilo, HI, USA and
Adjunct Faculty, Capella University, Minneapolis, MN, USA
SRahman@Hawaii.edu

## ABSTRACT

*As a result of the increased dependency on obtaining information and connecting each computer together for ease of access/communication, organizations risk being attacked and losing private information through breaches or insecure business activities. To help protect organizations and their assets, companies need to develop a strong understanding of the risks imposed on their company and the security solutions designed to prevent/minimize vulnerabilities. To reduce the impact threats have on a network, organizations need to: design a defense layer system that provides multiple instances of protection to prevent unauthorized access to core information, implement a strong network hardware/intrusion prevention system, and create all-inclusive network/security policies that detail user rules and company rights. In order to enhance the overall security of a basic infrastructure, this paper will provide a detailed look into gathering the organizational requirements, designing and implementing a secure physical network layout, and selecting the standards needed to prevent unauthorized access.*

## 1. INTRODUCTION

The continuous advances in information technology and security techniques provide individuals with more ways to access private networks and personal information. To this day, it is quite common to hear on the news about hackers breaking into companies/websites and stealing information. In fact, two of the most recognized hacking groups active today are Anonymous and Lulzsec. Anonymous is a group of hacktivists from all over the world that gain unauthorized access to websites/companies and advocate protests, in order to promote the freedom of information and the freedom of speech. In other words, they are trying to promote democracy, in their own way. The main goal of this group is to punish organizations for the way they act and for hiding information from the rest of the world. Along these same lines, Lulz Security (Lulzsec) attacked several high profile companies/organizations and leaked private information, exposing the need to secure networks and maintain safe user accounts. Due to real estate organizations becoming increasingly dependent on information technology systems to manage all of the property information for rent/sale, the list of clients and their personal information, and online solutions that will provide customers with the ability to obtain information/manage their accounts; there needs to be a company-wide understanding of the potential risks/legal issues that can occur from the basic network design and how to secure the network to prevent attacks. In addition, solutions will be identified in each section of the network structure, which can be implemented to protect information, prevent unauthorized access and reduce common risks. No one can guarantee complete security protection; however, taking a proactive approach in implementing a secure infrastructure, can be useful in preventing or minimizing internal and external attacks within a network. After this solution has been fully implemented, the company will need to re-evaluate the overall security plan/policies and make any modifications necessary, due to the constant evolution of risks/threats.

## 2. Background Study

Before a new security plan/structure can be designed or implemented, it is essential that we first look at what makes up the real estate organizational business model and the existing network/security model currently in place. By doing this, we will be able to determine: the strategic intent of the business, the requirements/needs that must be met, and the ability to expand or enhance the current network security structure to protect the network and increase data throughput. Even though data/network security is a common concern and no one wants to lose information, security is often overlooked and is the last step implemented within a network. The real estate business model being described in the following sections represents a general small business concept, which reveals real-life common security issues.

## 2.1 Organizational Structure

In simple terms, the real estate business model is the buying/selling or renting of property (which focuses on different areas of business such as commercial, residential, vacant lots, and etcetera), in order to earn a profit for the company and provide land to customers. The organization's model that we will focus on is to provide real estate property and services to customers at lower/more affordable prices. Today, with all the uncertainty of the economy and high unemployment rates, it is hard to obtain loans and/or purchase real estate through big lenders due to background/credit checks. In addition, a lot of people are afraid to make big purchases, in case they lose their job or are unable to make the full payments. Because these problems exist, the business model is designed to try and help people get back on their feet and rebuild their reputation/credit. However, to reduce business costs and overhead, the strategy used to achieve this mission is very limited and relies mostly on cheap forms of advertisement and word of mouth referrals. In general, the overall strategy used to accomplish the mission is to advertise their business by drawing in the customers directly that is actively looking for the product.

Although the above business model tends to focus on the lower priced items and waits for customers to contact them directly, the overall strength of the technique can be seen in the structure of the business itself and the expertise of each employee. The real estate company is divided into two major groups: the office personnel and the maintenance/construction crew. The office portion consists of several employees, including the: CEO, secretary, information technology specialist, finance specialist and the real estate lawyer; while the construction crew consists of the crew manager and crewmen. In the end, it is the human resources that provide the real estate company with the skills, knowledge and expertise needed to keep the business running/profiting. Part of the business strategy requires employees to research parcels of land to determine which are worth purchasing at auctions and fixing up, in order to rent or resell them at higher prices. The strategic intent "focuses the organization on key competitive targets and provides goals about which competencies to develop, what kinds of resources to harness, and what segments to concentrate on" [1]. Although this type of business could effectively be ran out of one office, some business models include several offices across town that either focus on different products within the market or makes it more convenient for their customers.

## 2.2 Network Requirements/Company Needs

Once the general business model has been defined, there is still another task that a specialist must perform before they can fully analyze the current network structure and suggest a new implementation plan – interview the CEO and the end users to determine what hardware/software requirements are needed. After they are collected, these customer requirements will be included within the new implementation plan and will contain service, organizational, and technical needs. Even though the employees may not know exactly what they need directly, they will be able to provide a list of applications or processes used, in order to perform their daily tasks. Gathering all of these processes together will help define the requirements needed by the customer, in order for them to perform their duties and to be satisfied. Although a general design could be immediately implemented that would satisfy approximately 80% of the company needs, there are still two main concepts that would not be satisfied: what changes have been made recently to the organization's daily process and have there been any changes in application or laws. In order to make sure every requirement/need is assessed and implemented into the network structure, the following document will outline the basic networking security standards needed in order to secure the business infrastructure. A little work with the organization's employees will be required to fully understand the daily processes/law, which will be used to satisfy the remainder of the customer requirements. The following benefits will be introduced by the full implementation of the new infrastructure security plan: the protection of the private corporate information, keeping the company in compliance with the Health Insurance Portability and Accountability Act (HIPAA) of 1996, the prevention of unauthorized access to the network, connecting all of the offices together, the implementation of remote access for the CEO/crew, and enhancements to expand the company.

There are several requirements and constraints to be considered, which can be broken down into the organizational and technical classifications. Some of the organizational needs of the real estate company include improving data access by creating a central database, expanding the company and connecting all locations to the same network, increasing staff awareness, reducing costs, increasing customer satisfaction and increasing website ease of use. Likewise, the technical requirements needed to make sure the business runs properly, include: increasing the network performance by upgrading/enhancing equipment, reducing bandwidth issues/dependency, implementing security, maintaining adaptability of the network, simplifying the network management to make it more user-friendly, and add mobility/QOS/voice solutions. In order to create a solution to accommodate these needs, there are few constraints in which the specialist will have to work with: limited computer knowledge among current employees, a limited budget, the need to create a single/multi vendor solution, and all of the hardware must work together and work with old/new equipment combined.

## 2.3 Existing Network Structure

In the general business model being discussed, the current functional design for the real estate network is a very simple structure with extremely limited security features. The network consists of multiple desktop computers connected directly to a DSL modem/hub and several laptop computers that are not connected to the Internet at all. Although the network has been set up to allow each desktop computer direct access to the Internet, the entire infrastructure security consists of one built-in firewall (within the Ethernet modem) and a free anti-virus application installed on the individual desktop computer. In addition to the above set up, in order to communicate with other employees or with another office, each employee must meet in-person or call the other individual directly on the phone. Communicating on the phone is efficient, however, the corporate phone lines are usually busy due to the new/existing clients calling in. As we can see from the real-life network diagram to the right, the overall design of this network model is simple and there is limited emphasis placed into security.

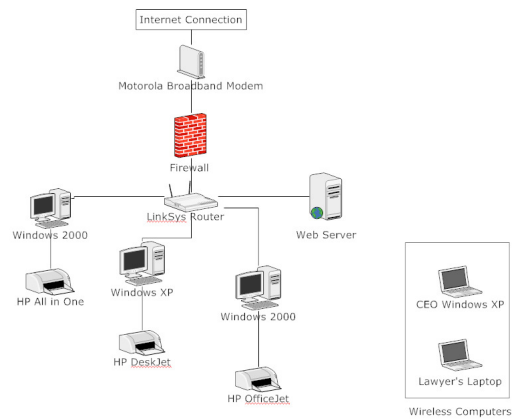

Real Estate Network Diagram

Figure 1. Network Diagram

## 3. Vulnerabilities and Methods of Protection

When running a business with an insecure network, such as the one in Figure 1, there are several types of threats that can cause harm to the organization/corporate information. By acknowledging the potential problems that may exist within the network, the administrator will be able to take the correct procedures to monitor traffic and/or prevent the problems from occurring.

## 3.1 Threats and Countermeasures

Potential threats can occur from both the inside and outside of the local network. Below are several network-oriented vulnerabilities, according to Zachary Wilson, that should be addressed within the security plan, to protect the organization and the sensitive customer information [2].

### 3.1.1 Intrusions

Intrusions represent the way people gain access into a system/network or access to the customer information found on the network. Examples of intrusion techniques can include physical intrusions (physical access to the computer/information, giving themselves rights), system intrusions (the ability to use an exploit to gain admin privileges), and remote intrusions. Physical intrusions can range from simple to complex methods, which allow users to gain access to the operating system and perform the intended activity or to give themselves administrative privileges to access specific areas/perform specific functionality. One example of a physical intrusion technique that bypasses the user account password within a Windows/Linux system without any effort is through an application called Konboot. Konboot is an open-source software application (paid version is 64-bit compatible) that is downloadable from the web, which allows a person to log into any pre-made user account without knowing the password. In order to do this, all a user needs to do is burn the application to a CD/flash drive and reboot the machine to the device. This boot process will then hook the BIOS and temporarily modify the kernel to allow a user to bypass the operating system (OS) user authentication during login. Although the login screen appears to be the same, the user can leave the password field blank or insert random text, and they will be given access to the OS (the Linux version is performed through the prompt and not a GUI). However, there are ways to protect a computer from Konboot. One method is to disable the ability to boot the system from the CDROM or USB drive and to password protect the BIOS area. This will help prevent unauthorized access/modification of the computer's BIOS and will prevent users from accessing administrative accounts. The second method is to encrypt your entire hard drive, using TrueCrypt or PGP, since Konboot cannot access encrypted folders or decipher specific algorithms. The last method is to get a BIOS protector application from the same company (most likely will need to be purchased), which will prevent any application (including Konboot) from booting [3].

### 3.1.2 Design Flaws

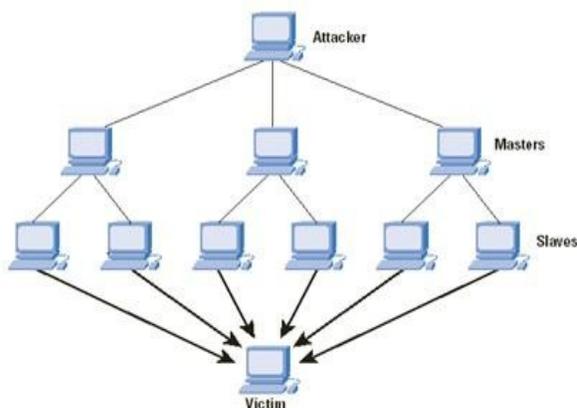

Figure 2. Master/Slave Architecture [4]

These include flaws within the overall structure of the network or equipment and issues created by a network administrator directly, which can consist of TCP/IP protocol flaws (examples include smurf attacks, ICMP Unreachable disconnects, IP spoofing, and SYN floods), man-in-the-middle attacks, session hijacking, and poor system administrator practices (empty or easy-to-guess root/administrator password). However, one of the most effective attacks and widely used methods today (due to the ease of use and lack of an easy fix), is the distributed denial of service (DDoS) attack. The DDoS attack is a method where the attacker employs botnets and/or zombie computers (innocent infected computers that join the attack unknowingly) to send packets of information to a specific target. Due to the large amount of data being sent to the target at once, the attack leaves the target inaccessible by flooding the network (consuming the entire bandwidth) or by crashing specific hardware/software. In order to obtain these zombie computers and maintain the connection, the attacker: scans the network/internet for vulnerable systems that have limited security features, exploit the machines through the use of a buffer overflow/code injection/rootkit, create a backdoor for future connections, and then update the security to prevent future attackers. The image on the left shows a typical botnet design structure, in which the attacker uses the compromised machines to attack a target.

### 3.1.3 Network Security

There are several major threats within the generic real estate network model that could cause a breach or even a loss of data. These threats include: the use of one server or desktop computer to store all of the information and to run the business website (they need to keep the private network isolated from the public network), the lack of a firewall (both perimeter security and a stateful security service) to protect the server and private data from intruders, and data integrity or privacy through encryption/SSL/VPN connections.

### 3.1.4 Operating System Security

Although network security is a big issue, several vulnerabilities exist at the software level as well. Some of these issues include: outdated operating systems for both workstations and the server, desktop computers/servers are not patched or up to date, no system documentation at all, and no user authentication to prevent access from outside sources.

### 3.1.5 Other Vulnerabilities

In addition to the above issues, other threats to the network can include clear-text sniffing, encrypted sniffing, keystroke logging, address resolution protocol (ARP) cache poisoning, the lack of training for the IT employee and the lack of verification that a change is implemented correctly. The purpose of ARP is to find the correct media access control (MAC) address of a computer on the network that matches a specific IP address. In other words, ARP maps the 32-bit Internet Protocol (IP) and the 48-bit MAC address within the data link layer. Since switch networks look at the data first, they only forward data packets to the intended recipients, based on the MAC address table. Because of this extra security, switched networks tend to be more secure and faster than typical networks. In order for an attacker to receive this data, they need to invoke a process called ARP cache poisoning. "ARP cache poisoning is the act of introducing a specious IP-to-Ethernet address mapping in another host's ARP cache. This results in diversion of traffic either to a different host on the LAN or no host at all" [5]. Eventually when the host's access resolution table is updated with the new information, it will begin sending all of the data to the attacker – who can then sniff the packets for information or modify the data being sent to another user. This process can be performed through an application called Cain & Abel. With a built in native network sniffer (Cain), the application is able to detect/record and crack passwords from a given LAN network in real-time (Abel).

Table 1. Impact Level of Several Problematic Areas

| Potential Threats | System Impact Level | | |
|---|---|---|---|
| | Low | Moderate | High |
| Private data is not isolated from public data | | | √ |
| No backup data or server available | | √ | |
| Lack of firewalls, routers, intrusion detection | | | √ |
| Patches and fixes | | | √ |
| Lack of encryption/secure connections | | | √ |
| Outdated Operating Systems | | | √ |
| Lack of system documentation | √ | | |
| User authentication | | | √ |
| QuickBooks Pro and ActiveX | √ | | |
| Custom-made program | | √ | |
| Training and certifications | | | √ |
| Verify changes before being implemented | √ | | |

### 3.1.6 Countermeasures

To help prevent these issues from occurring, there are several things that can be implemented and followed. The real estate organization needs to first start out by implementing a strong infrastructure and then securing this network through the implementation of layers/policies. Once the underlying structure is set in place, then all forms of transmissions (wired and wireless) should be encrypted or secured to prevent others from accessing the information. After the physical portion of the network has been secured, the next step is to train all of the employees on things they can and cannot do while on the network. During this training section, it will also include the need to create a strong authentication username/password, in order to limit access to corporate computers and to keep track of each employee if needed. It is up to the administrator to enforce that each employee understands that passwords cannot be documented anywhere and the security risks behind it. The last major step for the company is to comply with all of the requirements/standards and to constantly monitor for anything out of the ordinary (on or off the network). Table 2, below, shows a chart comparing specific security threats with areas of operation and reveals the actions necessary to prevent the security threats from occurring [6].

Table 2. Addressing the Defects

| Security Dimensions | Security Threats | | | |
|---|---|---|---|---|
| | Interruption | Interception | Modification | Fabrication |
| Access Management | | Control access to network elements and links to prevent wiretapping. | | Protect against unauthorized use of network resources. |
| Authentication | | | | Ensure the identities of communicating parties. |
| Non-repudiation | | | | Detect and provide evidence of forgeries on the network. |
| Data Confidentiality | | Encrypt data so that it is not meaningful to attackers. | | |
| Communication Security | | Prevent data diversion. Ensure data only flows between designated endpoints. | | |
| Integrity | | | Prevent and detect unauthorized creation, modification, and deletion of data. | Detect forgeries on the network. |
| Availability | Ensure network elements, services, and applications are available to authorized users. | | | |

| Security Dimensions | Security Threats | | | |
|---|---|---|---|---|
| | Interruption | Interception | Modification | Fabrication |
| Privacy | | Prevent unauthorized parties from obtaining private information of network use. | | |

## 3.2 Information Security Governance

Since the real estate agency creates/saves a lot of personal identifiable information that is considered valuable, it needs to develop a method to keep this information protected and prevent unauthorized access. "Governance is the set of responsibilities and practices exercised by the board and executive management with the goal of providing strategic direction, ensuring that objectives are achieved, ascertaining that risks are managed appropriately and verifying that the enterprise's resources are used responsibly" [7]. In other words, information governance is the set of tools, people, and business procedures used to ensure that security is performed within the organization to verify requirements are being met. Security is the responsibility of everyone in the organization, where everyone must work together to ensure protection of the data/business and that all of the business components work well together.

## 3.3 Control Mechanisms

To help ensure the protection of client information, information security control mechanisms are put in place. Information security control mechanisms are the individual controls implemented within a system to prevent unauthorized people from gaining access to information and to protect the overall confidentiality, integrity, and availability of information. In order to prevent people from accessing the information found on the network, four main types of controls need to be implemented: preventive controls (firewalls/lists/policies) to block unauthorized access, detective controls (surveillance/logs/audits) to identify and characterize incidents, corrective controls (IPS or previous responses) to fix the issues and get everyone back to work, and recovery controls (backup systems/power supplies/spare parts) to restore services and resume business [8]. As mentioned before, the best defense within an organization is to setup a layer system of security devices. A network with only one security layer is extremely vulnerable and once this single defense layer has been broken into, there is nothing left that will prevent the unauthorized user from causing harm, gaining permissions, or doing anything they want to. To help combat this problem, an organization should implement several layers of protection to help catch/delay any users that may bypass the outer perimeter. In order to keep the network secure, the company will also need to implement: Secure Socket Layers, antivirus programs network-wide, firewalls, remote access with VPN and login, secure wireless internet, DMZ structure, policies, backups, and redundancy. In addition to the above network needs, any business that contains information onsite also needs to implement physical preventive controls such as: having doors installed throughout the building, integrating door locks (especially on perimeter and server room doors), installing and using a security alarm, and leaving on an external light near the building. The real estate company will also be required to implement internal methods or documentation used to help mitigate issues, including: user authentication or an access control list, policies and procedures, laws, and of course training on all of the material.

## 3.4 Defense Layers

As briefly mentioned throughout the beginning of the paper, the most secure method of protecting a network is to implement a layer system. This method provides multiple instances of protection, in order to prevent unauthorized access to an organization's core information. If one of the layers is breached, then the next security layer will catch the unauthorized user; giving the IT specialist more time to detect the intruder. When developing a layer system for your network, the four key layers are the: perimeter defense, operating systems and server's protection, host protection, and information protection [9].

As the name implies, the perimeter defense of a network represents the hardware/software set up, which is used as the first line of defense to protect the network from attacks. This layer usually consists of firewalls, routers and proxy servers. In order to prevent attacks at this level, the administrator should configure/test the proper filters for the network and implement an anti-virus/intrusion detection system. The second layer, operating systems and server's protection, is the ability to secure each computer/server operating system in order to prevent attacks on the network. These forms of protection can simply include physical location and locks, authorized or restricted access, patches, service packs, or even hotfixes. After looking at the previous layer, we continue hardening the system by implementing security on the internal workstation. According to Karnail Singh, there are several important security concepts to keep in mind when working with workstations: create and implement a user access policy, limit network resources, install firewall security/antivirus software, backup information, do not install modems, enable all logging and discard any unused media to protect sensitive data [9]. The last layer that should be implemented within a network is the data/information protection layer. This layer represents the steps necessary for an administrator to implement a secure environment and to reduce as much risk as possible. Although there are several pieces to this section, the main focus is to implement a strong operating system that has many security features, data/network encryption to prevent outsiders from gaining access to information, strong and unique passwords, and sensitive storage protection methods.

Although infrastructure security can begin by implementing a single tool or application, the entire network relies on the success of that tool to block all unauthorized access and risks. The multi-layered strategy approach is designed to include fail-safes within the network, to help prevent the likelihood of unauthorized access to the corporate information or a full network breach. In order to improve the original real estate infrastructure and to make it more secure, the concept of defense layers will need to be integrated.

## 4. NETWORK DESIGN AND FRAMEWORK

After obtaining all of the requirements that need to be included within the network and understanding the layered structure of the network/business, the next step is to begin developing the infrastructure from the ground up. We will first look into the overall structure of the network and the locations where all of the switches/servers are located. By going into this much detail, we are able to touch upon methods of: speeding up transmissions, Quality of Service (QoS) priorities, and the use of specific protocols. Following these details, this paper discusses the implementation of both wireless networks (secure and unsecure) within the office. In addition, we will then expand our views and see the effects of the company's physical location, training programs, auditing, and database management techniques.

### 4.1 Improved Network Design

Startup businesses like to deploy single brand solutions to increase interoperability between network areas and for network-wide customer service, even though it may not always be a requirement or the most appropriate method to use. This method makes it very simple to request support (if needed) and helps decrease the risk of network modules not working together well. Although the Cisco Enterprise Architecture is too elaborate and the price is too high for a smaller real estate organization to afford, specific aspects of the overall design and model structure can be applied to the improved infrastructure. The Cisco Enterprise Architecture is designed to implement a functional modular network, which consists of the following six modules: Enterprise Campus Module, Enterprise Edge Module, Enterprise WAN Module, Enterprise Data Center Module, Enterprise Branch Module, and Enterprise Teleworker Module [10].

The first functional area of the Cisco Enterprise Architecture is the Enterprise Campus. The main purpose of this module is to incorporate all of the servers and data into one location to maintain security, authentication, and to prevent information from being spread out. By combining all the information for both offices into a single location, the transfer speed is much faster between servers and it allows the corporation to maintain better security on the servers/databases. Although the second office can do light processing at its own location, it will need the ability to connect to the main office, in order to access corporate information/databases. Within this module is the addition of a server farm, which incorporates the internal networking aspects, such as internal communications, email, a file server for internal use, and the DNS server [10].

The second functional area consists of the Enterprise Edge module. In simpler terms, this is how each network is set up to communicate with: itself, the other office, the customers, and the rest of the Internet. Not only does this module consist of the physical connection, but it also includes the firewall put in place to protect the network from outside sources and the equipment used to help reduce any issues (router compared to a hub). One issue that may cause severe problems for the company is if their current Internet Service Provider (ISP) goes down. To prevent this from being an issue, a second ISP should be installed as a backup or to work in tangent. Duplication of areas will be mentioned in more detail within the redundancy section, but if a lot of data or business applications are required to be used on the web, the business cannot afford for the Internet service to go down. As of right now, both offices will have a DSL connection to the Internet and the ISP will remain the same with the VPN function enabled. The next part that needs to be implemented is the Wide Area Network (WAN). A WAN basically allows two separate local networks to communicate as one whole network and eases the ability to share files or access between the two buildings/networks. The main office building has a faster connection to the Internet and contains a Virtual Private Network (VPN) server. A VPN is a secure connection formed that allows one computer to talk with another computer, anywhere in the world. Since the remote office also has a fast connection to the Internet, there are two choices to connect the remote site to the main site. The network can be designed to direct the computers in the remote site straight to the server itself via the Internet or we can create a VPN tunnel to keep the information secure. A VPN tunnel is basically an encrypted connection between the host and the end user, which allows information to be passed back and forth securely. However, in order to do this, the remote office will need to install VPN client software on a workstation to connect to the corporate VPN. The Enterprise Edge module also contains a network of servers for the actual web portion of the business (database server containing tenant information, web server, and the ecommerce server), since it uses the availability of the server farm and the connectivity of the Internet module. The entire web portion is being implemented in order for the company to expand their business into the digital realm in the near future and to provide their clients with several additional features [10].

The next functional area is the Remote Enterprise Module. The first module defined in this area is the Enterprise Branch module. The second office has their own local network where they can communicate with each other, as well as, an Internet connection and VPN software to communicate with other networks. This allows them to connect to the main office and have the same security that the main office has. The second module defined in this area is the Enterprise Data Center module. To incorporate this into the network design, database and file servers have been added to the network infrastructure. The database server will store and host all of the client information for the company, via their corporate website. This will provide clients access to their information and will eventually allow them to make payments from their own home. If the client has direct access to information, however, an authentication server and a secure connection will need to be implemented on the client end. The file server will host internal corporate files and client/tenant information, to keep track of all the financial and legal information. The last module defined in this section is the Enterprise Teleworker, which is represented by the VPN client software and VOIP telephone system allowing the remote office to connect to the main building [10].

There are several services that would be good to add to the real estate network in order to protect it, but only a few are selected in order to reduce the initial costs. Another reason for not implementing all of the services, as mentioned above, is that the overall company size is still very small for the time being. As they expand, so will their network and the need to add more functionality. To help secure this network, all traffic will initially have to go through a router and firewall to help reduce unwanted or harmful traffic. One way the traffic will be filtered is through the hardening or closing of network ports. Eventually, an authentication server will be added to the network to only allow staff access to the network and the customer database portion, when clients/tenants log in from the web. Since a handful of clients bring their laptops in with them for business reasons or use them during their waiting times, an unsecure wireless access point before the firewall has been installed, to allow customers the ability to access the Internet. This is in addition to the secure wireless access point on the network, which is installed for corporate users only.

When creating a network, the most important thing is the ability to direct traffic to the desired location. In order to ensure this can happen, administrators add redundancy throughout the network to provide alternative routes for traffic to be transmitted, to help manage load balancing and network availability [10]. Load balancing is simply navigating data to different paths to keep the transmission bandwidth equal. If one path is running slow or has too much data, it is designed to navigate down a different path to decrease the transmission time. Since the real estate corporation needs to be connected to the web at all times (the business aspect needs to be constantly hosted, clients are allowed request work or access their own information from the web, and the remote office needs to be able to connect to the main office), they can not have the Internet access go down. By adding a second ISP or backup solution, the company will be able to still access the Internet while repairing or congesting the first one. However, if another ISP is added to the network, another router and possibly firewall will also have to be added to the network in order to support the traffic. The router is a must and will be able to connect to either switch or firewall, depending on how this is set up. If a router is down, the traffic can still be transferred to the rest of the network. Even though all of the data can be routed through one firewall, it might be a good thing to have two just in case. Since the firewall is needed to protect the entire network, the business cannot afford for this to go down and leave the network wide open. The last bit of redundancy that is needed is another switch accessing the servers or work computers. Again, if the switch to the servers goes out, there will be no way to access the corporate information or web server information. Therefore, by adding a second path to the servers, this allows traffic to still be able to access the internal servers in case an issue arises. For the case of the work computers, this allows the Local Area Network (LAN) access to the corporate information as well.

When performing an analysis of the existing network, the following tools will be used to retrieve information on the overall hardware status: Simple Network Management Protocol (SNMP) and Management Information Base (MIB). In order to verify the information, the network monitoring application would need to be installed on one of the workstations and then the SNMP tools will need to be installed/enabled on each of the main hardware devices (routers, servers, etc.). In order to read the information detected from each of these areas, the administrator will need to have an MIB set up. An MIB is a database that stores the information collected from the SNMPs and then makes them viewable/readable for the administrators. These MIBs need to be installed, configured, and then registered. From this point on, in order to view these MIB files and to check on the status of the current network, the administrator will need to use an application such as MIB Browser, WhatsUP, and etcetera. Some of the information that will be useful in checking the status of the network, is: the IP address and port number being used (to make sure they are not blocked), the number of active threads and processes (make sure data is being worked on and sent in/out), and even the on/off status. This will display whether each section is working correctly or not, in order to see if information is being passed through or if it is not being received at all. Remote Network Monitoring (RMON) is an extension of SNMP that is designed to look at MAC layer data and help monitor the traffic activity on a given network. After enabling the SNMP and adding a dedicated RMON probe to a router and switch, the RMON tool will need to be enabled. This tool is designed to capture the packets being sent through the device and the information/statistics is saved within the RMON. This system will be set up to alert the administrator if the network traffic goes over a defined threshold, report if errors occur, and will provide packet pattern information (number of packets, size, network utilization, errors, collisions, and statistics for the hosts). RMON allows statistics to be collected on a network, even if a failure occurs between the probe and the console [10].

## 4.2 Switch/Server Locations

The core layer, or backbone, represents the foundation of the network. This layer is designed to transport a large amount of data, quickly and reliably to the remaining sections of the network. The reason it is called the core layer is "because all other layers rely upon it...its purpose is to reduce the latency time in the delivery of packets" [11]. The best place to put the core layer switch in the main office building is above the server farm. This is the most critical spot, which requires a lot of information to be passed through accurately. By placing a high-speed switch here, it should switch information quickly and allow better communication throughout the rest of the corporate network. Not to mention, this is one of the most critical areas of the network, since it keeps the company network running. The remote office has a slightly different approach since it does not need all three layers. This office will only consist of a distribution layer (to distribute the connection to/from the internet) and an access layer. In the main office, two more switches within the distribution layer are added to help distribute the information being provided by the core server.

The last set of switches added to the main network belongs to the access layer. The access layer "controls user access to network resources" [12] and is designed to connect end users (or workstations) to the network. Since the access layer generally connects a workstation, hub, or database server to the network; an access layer switch has been placed within the following areas for the main office network: the switch connecting to the database/web servers, the switch connected to the internet, and the switch that is connected to the end user VLANs (see Figure 4 on page 16). As for the remote office building, two access layer switches have been included within the network. One switch connects the distribution switch to the first VLAN and the second switch connects the distribution switch to the second VLAN.

All of the main corporate servers for the network are located on their own network, known as the server farm. Since this data is critical for the corporation, it is placed at the end of the network behind the core switch (which includes a firewall) and the network firewall, for safety. Even though a fiber optic connection from the server farm to the main switch would be faster, it would be too expensive for the real estate company and would be a complete overkill. The CEO does not want to waste the money and would not see the real advantage of a fiber optic cable, therefore a typical category 6 (cat6) cable would be fast enough for the entire network to use. Using cat6 cabling guarantees a transmission speed of 1 Gbps or more, which has a higher bandwidth than the cat5 cable. This allows the network to handle faster speeds and better support for streaming video/communications with the clients or remote office. In order to reduce some of the cost of the network and knowing the company would not see the benefit of the extra cost, the server farm will also be connected to the core switch using a cat6 cable.

When it comes to the core switch, the company should install either a Cisco Catalyst 5500 series or a 4500 series. Unfortunately, the Cisco Catalyst 5500 switch is no longer being made. Therefore, finding replacement parts and upgrades can be rather difficult. Although the 6500 series is a great product and it provides a lot more functionality, it is a complete overkill for a small to mid-sized business and would not be cost effective. A good core layer switch for a small to mid-sized campus setup is the Cisco Catalyst 4500 E-series. This switch is designed to transport data, video, and voice securely and quickly. Some of the hardware-based functions include: "quality of service, multicast, security, IPv6 and integrated management features" [14]. The best part about this switch is that it is backwards and future compatible, especially since IPv4 addressing has ran out (been declared dead) since February 3, 2011. This makes it easier to upgrade and allows the company to keep up with some of the newer technologies. In order to connect the branch office with the main office, the real estate company will have to purchase and deploy a 2800 series router in the remote office, as well as, the main office. Although the company could use a cheaper model, this model can provide multiple connections and a highly secure connection when providing video, voice, and data services. This will allow the company easier and safer access to distribute the 100 mb images, VOIP telephony, and also videoconferencing. The 2800 series router also includes T1 ports for communications with the corporate office, Ethernet connectors for Internet access or forwarding, IP phone users, and VPN tunnels [15].

| Priority Level | Traffic Type |
|---|---|
| 0 | Best Effort |
| 1 | Background |
| 2 | Standard (Spare) |
| 3 | Excellent Load (Business Critical) |
| 4 | Controlled Load (Streaming Multimedia) |
| 5 | Voice and Video (Interactive Media and Voice) [Less than 100ms latency and jitter] |
| 6 | Layer 3 Network Control Reserved Traffic [Less than 10ms latency and jitter] |
| 7 | Layer 2 Network Control Reserved Traffic [Lowest latency and jitter] |

Figure 3. QoS Priority Levels [13]

To help maintain the traffic across the network and between office buildings, I will also have to include QoS. The QoS priorities list will have to include: IP routing, voice, interactive/streaming video, traffic shaping, and even traffic policing. This will ensure the proper traffic is flowing correctly and if a problem occurs, the system will deliver the traffic according to the priority table above (Figure 3).

To ensure that the traffic flows efficiently throughout the network and in between two separate networks, the best interior gateway protocol (IGP) and an exterior gateway protocol (EGP) needs to be selected. Example protocols used within the IGP include: Routing Information Protocol (RIP), Open Shortest Path First (OSPF), Integrated Intermediate System-to-Intermediate System (IS-IS), and Enhanced Interior Gateway Routing Protocol (EOGRP); while the exterior gateway protocol mainly consists of: the Border Gateway Protocol (BGP), which is used to connect two autonomous systems together [16].

The OSPF protocol is a classless routing protocol and can be used within all three layers of the Enterprise Campus: the Building Distribution Layer, the Building Access Layer, and the Enterprise Edge. This protocol should be implemented within the real estate company, since it supports multiple vendor devices. Currently, the real estate company has a running network and all links on the network connect correctly. Although the hardware setup is rather minimum, they currently have existing hardware and cannot afford to replace every single item in the beginning stages of the upgrade. Because of this, OSPF supports multi-vendor devices, where other protocols such as EIGRP do not. Another reason why multi-vendor support is worthwhile is in case hardware needs changed. In order to increase functionality across the network and keep the overall costs lower, the main switches will be left as Cisco based and the smaller routers will be a non-Cisco based solution. This will help keep costs lower and add more functionality to the network. In addition, if the company decides to ever upgrade the switches/routers, they will be able to install a hardware device from any vendor and it should work correctly. This gets rid of the necessity to purchase the same vendor, saves on time (no need to reconfigure the actual data part), and makes it easier for the company to replace a device without having to know exact details since their network/IT knowledge is limited. One of the major problems with having a single vendor solution for a network is that a company is tied to only the vendor's technology. If the vendor grows, then the company can grow along with it; if the vendor stops growing, then the company must also stop growing or upgrade to a new vendor (which can be expensive). By building a company around one vendor, a company relies on their products and is putting all of their eggs in one basket. Not every company makes the best of everything; they usually specialize in a product and then create other products that work well together and keep the customer buying from the same company. Since this is the case, confining a company to one vendor means they might not get the best quality in another area of the network.

Another advantage to using the OSPF protocol is when a host obtains a change to a routing table or detects a change in the actual network, it immediately multicasts the information to all hosts within the network in order for all of them to contain the same routing table information [17]. Other protocols tend to use a higher CPU or memory usage, since some protocols send a message every 30 seconds. This message again contains information on what all has changed within the network and to update the routing table for all other hosts. Since OSPF only sends the information when something has changed, the strain and processing used on the switch/router remains lower until it needs to determine the change/best route to use. As you can see, using the OSPF protocol within the LAN for both the main office and the branch office can provide many advantages to help information pass from one location to another. Other advantages that the OSPF protocol will provide the real estate company, include: fast convergence times, very good path utilization, the shortest paths are calculated well, authenticated users/routing updates, and OSPF "supports VLSM to achieve better use of IP address space" [16]. Besides being one of the most commonly used protocols used today, OSPF is an efficient protocol. Even though it may be a little harder to use or configure, the protocol can support a large network and it can calculate the shortest path to send information. As for the exterior WAN connection, a point-to-point link from the Internet Service Provider (ISP) to the network will be set up. But in order for the network to help differentiate between its own traffic and the traffic from the ISP, the Border Gateway Protocol (BGP) version 4 will be implemented.

## 4.3 Wireless Network

Networks mainly consist of wired connections, however, wireless solutions have started to make a huge impact and are being installed in all areas of the network. There are many advantages to implementing a wireless network or access point, which includes: mobility, range, lack of physical wires, and etcetera. By implementing this into the overall network plan, it is even possible to pass these benefits on to the customers by integrating two access points per building, in order to accommodate the customers and employees.

Before a wireless network can be implemented within the office, the network administrator needs to perform risk management and identify the problems associated with the business. Risk management is the ability to identify, assess and prioritize risks in order to reduce or remove the impact they have on the overall project. "The principal goal of an organization's risk management process should be to protect the organization and its ability to perform their mission" [18]. On the other hand, risk assessment is the process of determining what information needs to remain secure and to document the specific events that can cause a loss of data or a breach of confidentiality. Project risks can be identified through the use of historical data of previous projects or experiences, common risks associated to most projects, and even unique risks found specifically for the project's functionality. Some of the risks to the wireless network can be broken down into the following categories: accidents (natural events, hardware failure), malicious software (viruses/worms/spyware), external threats (thieves, physical/electronical attackers, employees from other companies), and internal threats (employees release confidential information, break equipment, download items that contain malicious software, break any policies or user agreements).

By allowing employees to work with private customer information, each one will need to sign a non-disclosure agreement to prevent any information from being leaked and they will also be prevented from taking documentation home. If an employee decides to take any work papers or information home with them, they need to get each paper double checked by a colleague prior to leaving to ensure no private information is contained on the paperwork. This will prevent private information from exiting the building and being used in other locations.

## 4.4 Physical Layout

To determine the best wireless standard and access points to implement, we need to focus on the overall layout of the real estate office. Between the physical size/shape of the building and the layout of each of the employees, we will need to determine the speed required to transmit the data to them and the strength of the signal required to connect to each computer. The real estate office is located in a small office building on the outside edge of a major city (low to medium interference), in which it consists of one floor in the shape of a rectangle. According to John Brandon from Laptop magazine, a typical wireless-n router can send a signal to approximately 1,000 feet. However, the speed of the network starts to degrade when the device is in the 300-600 foot range or if other wireless networks are present [19]. Since the router has a very good range overall and the office building is relatively small, the real estate company will be fine implementing a wireless 802.11n router to provide the wireless Internet access. By selecting the wireless-n router, the company is: capable of lowering the signal strength anytime they want to reduce the network from being spotted or used (lowering the service set identifier), increase the internal/external transfer speed, and using newer technology to prepare them for future needs/capabilities. As mentioned earlier, due to the security level being implemented on the router, there will be two access points being installed within each of the real estate offices: one public access point near the front of each office for the customers and one secure private access point in the back office for the employees.

The reason for two different access points is to provide a public connection for the company's clients to connect to (hotspot) and a private connection that only the employees can use. The public wireless access point will allow the real estate customers to connect to the Internet in order to surf while waiting, check their property information, or view their checking account balance to verify they can go through with the business transaction. Some Internet websites will be filtered/blocked, but this will provide a service to the customers. The public wireless router will not have any encryption installed (although they will be warned), in an effort to help make it more convenient for clients to access the Internet and quicker for them to access pages. However, since there are a few security issues with this method, a few precautions will need to be made. Since the network switches used within the network support virtual local area networks (VLAN), both wireless access points will be placed on a separate VLAN. This acts as though each wireless access point is on a separate network, in which it will have its own SSID, access permission, and security settings. The major advantage to this method is keeping customers in the public VLAN and allowing only employees access to the private network. Customers will remain confined to the public network only and will not be able to explore the entire real estate network. One thing that can be defined within the security and access settings is which users are allowed access to each network. By changing these settings, a user will only be able to access one network and will not be able to log into the other one. This will prevent customers from gaining access to the secure network and will also prevent the employees from transmitting sensitive data across unsecure lines. Another security item that is important to include is a parental control policy for the public connection. The network administrator will need to define what type of websites can and cannot be accessed (race, adult material, streaming video, and etcetera), as well as, limit the amount of bandwidth usage. By limiting the bandwidth usage, they will be able to make sure that the business portion of the Internet can still function normally and the customers will not "eat up" the speed or transmission space needed to function properly.

Since the customer access point is wide open, it is possible that several people may log into the network at once. In order to help protect each client from another user, I will need to set up client isolation within the router. Client isolation allows a client to access the Internet, without them having to worry about other clients snooping on them or their hardware. Without client isolation, other clients would be able to identify their IP Address and connect to/explore their computer system. Even though the router helps clients connect to the Internet safer, the client is still left wide open once they are connected to the actual Internet. By using this method, there are no extra software or passwords required to access the Internet or to use client isolation.

The private wireless access point is different and will require more security. As this paper has already alluded to, the wireless access point is on a different VLAN network in order to help separate the corporate network from the open network and to help keep it secure. According to Diane Teare, there are three forms of wide local area network (WLAN) security: authentication, encryption, and intrusion detection/intrusion protection [16]. To help secure the private network, the corporation will need to implement Wi-Fi Protected Access (WPA) with Advanced Encryption Standard (AES) with at least a minimum of 128-bit encryption. The implementation of 802.11i with AES is also known as WPA2. The encryption will need to be end-to-end and must use counter mode with cipher block chaining message authentication code protocol (CCMP) or another IEEE or NIST-approved key exchange algorithm for increased security. This will help improve the security across the network and prevent attackers from accessing the network. Even though AES is more CPU-intensive, it provides stronger security and helps prevent common wired equivalent privacy (WEP) vulnerabilities. Today, WEP authentication or passwords can be cracked within ten minutes, using a program called Kismac for the Macintosh (Kismet for the PC). In order to crack the WEP encryption key, the program detects wireless networks through a passive scanning technique (to prevent active probe requests) and needs to receive approximately 130,000 unique IV data packets. In order to create these packets, the user is required to perform an Authentication Flood attack and a Packet Reinjection attack. A Packet Reinjection attack is where the application sends data packets to the target network until the target network responds back. After the application receives a response back from the target network, the application injects itself into the network "conversation" in order to generate a large amount of traffic/packets. When all of the data packets are received, the user will perform a Weak Scheduling attack to recover the WEP encryption key. Because of the simplicity of this technique, WPA encryption or higher is recommended.

Since there are only a few employees that will need to access this wireless network within each office building, the business will be able to enhance the security of this network by also enabling MAC address filtering as well. MAC address filtering allows the network administrator to specify the only computers allowed to connect to the private network, by allowing specific MAC addresses access. Any other machine that attempts to access this network that is not in the list, will be denied access. Although this is a good way to keep most people out of the network, as in most situations, this only keeps the common individual out. To bypass the MAC address filtering technique, there are several methods the user can do which range in complexity. The simplest method requires the user to download two programs: a network sniffer and a mac address changer/spoofer. The user scans the target network for data being sent across and saves the packet information received. The packet itself contains the MAC address of the machine that transmitted the data and of course, the rest of the information. From this point on, all the user has to do is plug this MAC address into the spoofer and change their MAC address to mimic the one in the packet. Once the original device is removed from the network, the user will be given full access. Beyond this method, there are additional ways to bypass MAC address filtering that is more advanced and require the use of re-injecting packets, but provide the benefit of instant access to the network.

In case one of the networks should fail, a backup strategy should be implemented in order to help users obtain an Internet connection and retrieve any information that they may need. If the private network fails and employees are unable to connect to the Internet, the network administrator will be able to adjust the user access credentials and will allow them to connect to the public wireless access point. Although they will need to make sure confidential information is not passed over the public connection, they will still be able to connect to the Internet in order to perform their job duties. As for the public network, customers do not have to have a connection to the Internet. This is provided as a free service and the ability for a customer to gain access while waiting for their turn. If this network fails, then the customers will not be able to use the public access point and will be required to either wait until they leave the office or is given permission by the CEO to use a company machine (subject to monitoring). To help minimize any network from failing, two ISP connections will be available to help maintain a healthy Internet connection. If a specific VLAN network is to fail, the network will be fixed as soon as possible.

The corporation will also implement wireless monitoring and protection devices to maintain a secure network and prevent attackers from gaining access. In addition to the protection of network monitoring, the administrator will also need to implement an intrusion detection system. This will alert them if there is unusual activity or someone tampering with data in which they should not.

## 4.5 Training

The security of a network is only as good as each employee makes it out to be. To help maintain a secure network, the real estate company will need to hold group-training sessions for each employee. Training will cover: corporate policies, the importance of security to the corporation, meet and greet with the IT department so they know who to communicate issues with, the usage/protection of passwords and PINs, laptop and desktop computer security, virus/malicious software prevention and actions to take, safe internet browsing, and the consequences of not complying with corporate policies or requirements.

## 4.6 Security Auditing

Network security checks should be performed multiple times a year, in order to help detect any security flaws that already exist or that are created through other network changes. These network checks will be performed in-house, as well as, through an independent organization. After the network security is in place and the in-house audit is in progress, the network administrator will use several networking tools to help detect any vulnerability within the network. Some of the tools that will be used consist of: wireless LAN discovery and wireless key cracking, password capture & decryption, ARP or DNS spoofing (may need to adjust rules on the router), network management, wireless protocol analyzers, port scanning and even OS exploits [20]. Even though the corporation may be able to detect and repair a lot of security flaws themselves, hiring an independent organization will provide fresh ideas and experience/training to detect unseen security issues.

## 4.7 Database

When developing and deploying a database, there are several things that need to be considered in order to keep the information secure. In general, database security is the prevention of unauthorized user access and the prevention of unauthorized disclosure/alteration of the data within it. In order for the database to remain effective and within compliance of the law, the database needs to be available for the clients to use and all information needs to remain confidential. The methods below should be considered when implementing security within a database structure.

The administrator should set up the network to keep the database secure and out of reach from most users; which generally consists of keeping the database and all web functionality behind a firewall. After the network has been designed correctly, the administrator needs to look into the use of access controls and roles. Access controls assign specific views and privileges to limit users on the information that they can access within the database table. To increase security even more, the database administrator should also implement role-based authentication. Role-based authentication defines roles as various job functions, which each contain specific privileges or access rights. Users are then assigned to different roles (statically or dynamically) according to the responsibilities needed. A role-based privilege allows the administrator to make one change to the role and every user contained within it, automatically adjusts accordingly. This is a simple way to maintain user privileges and to make sure everyone within a certain department or status is treated equally. Another key factor is the overall value of the data itself, which includes data integrity and the implementation of encryption. Data integrity is ensuring that the data contained within the database is valid and has value. If the data is sensitive in nature, then the administrator needs to decide: the type of encryption method used to secure it and if it needs to be encrypted during transmission. In order to maintain a secure environment, it is also very important to maintain change control. Change control is the ability to track and regulate any changes made to the database or contents of the database. The last major necessity is to keep up to date on all security patches and disable any server functions that are not being used.

In order to keep all of the information secure, the network administrator will need to install a stable database to store the information. The real estate information will be contained on an Oracle database, hosted on a Unix machine due to the overall stability of the operating system and the ease of permission changing.

After discussing the overall changes made to the infrastructure and database areas within the sections above, the overall network design (below) has been modified to indicate these changes. As you can see, the new network structure remains expandable allowing the company to continue adding more devices at a later date and also includes redundancy to ensure that information can be accessed.

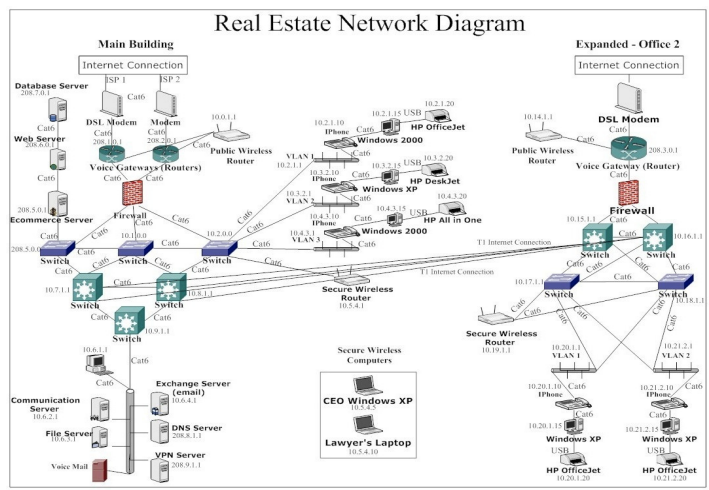

Figure 4. Expanded Network Diagram

## 5. ETHICS AND POLICIES

In order to maintain a healthy system and to make sure all of the employees are behaving correctly/securely, the last major step is to explain the role that ethics/law has within a business and to define the policies required for each employee to follow. By following these policies, the network administrator can reduce risk to the network and define a course of action in case an acident occurs.

## 5.1 Ethics and Law

There are ethical and legal implications that are related to running a commercial business and to keep customer information confidential. Ethics can be observed by the way the company: treats their customers and the approach used to gain their business, methods of securing the network to "keep the honest man honest", the ability for the company to advertise the standards used to secure their clients information, and the prevention of employees using/leaking the confidential information for their own good. The company is responsible to ensure all of the requirements and regulations have been met, it will adhere to the strict guidelines created, and to report any issues or conflicts immediately to get resolved.

But due to the nature of the content being stored within the company's database (personal and financial records), there are legal implications that relate to the network as well. The security system must remain stable and running at all times, to protect the corporate and personal information. In addition, the company must meet all regulations, in order to reduce the risks involved by both parties. If data loss occurs, the company needs to have a disaster recovery/incident response plan set up along with safeguards to clean up the mess.

Business entities have obligations to develop/use best practices when securing personal client information. There are federal and state laws that mandate processes for the secure storage, maintenance and transit of data (e.g., Health Insurance Portability and Accountability Act (HIPAA), Gramm Leach Bliley Act, and California Senate Bill 1386). Since the company deals with the buying/selling of real estate property, the Payment Card Industry Data Security Standard (PCI DSS) provides computer security requirements to improve payment account security. It applies to all businesses that handle credit card data and sets the standards for secure practices, including the use of: firewalls, antivirus software, security audits, network monitoring and more. In addition, it also requires the company to have an incident response plan in place.

## 5.2 Policies

In order to protect the company from disasters or loss of data, the network administrator needs to evaluate what material is important to the company and set specific guidelines for the network users to abide by, in order to maintain the safety of the network. All of these rules combined, make up the network security policy. The main purpose of this policy is to secure the company network, as well as, reduce risks/losses to important information. According to Diane Teare, the network security policy is made up of four areas: Network Access Control Policy, Acceptable Use of Network, Security Management Policy, and Incident Handling Policy [16]. The main purpose of creating the network/security policy is to provide a list of detailed information to alert others on proper etiquette or handling procedures. By documenting everything in a single location, it remains easily accessible for people to read and provides a specific outline that can be followed. In general, the above policies focus on the following functionality:

- Defines the roles and responsibilities of each of the employees of the business.
- Describes how violations are reported and the actions (consequences) taken when one occurs.
- Identifies who makes up the incident handling team to plan and respond to any network issue or problem that occurs. As well as, defines the logical step-by-step approach to identify, contain, and eliminate the threat.
- Infrastructure analysis of existing technology and equipment to identify the weaknesses of devices installed.
- Assessment of damage in the event that an accident occurs and the repercussions it has on the network.

- To help prevent the incident from reoccurring again, by creating/updating existing policies to: determine if a step was missing (which caused the incident), if the policy needs redesigned, or if it requires additional steps.

## 5.3 Network Access Control Policy

The first step when developing a network policy is to identify what information is important within the network and categorize them based on severity. This will help identify the level of security needed and develop a policy that will define which users are allowed access within the network. It is very beneficial for the real estate company to have an online connection, for many reasons: it is easier for customers to view the property available and the cost, allow government officials to view the material for sale/sold by the company, displays other services available through the company, and will soon provide customers with an interactive method to pay bills or view their own information. Because of these changes and improvements, the Internet can be a very risky place.

The network administrator needs to decide what applications are or are not allowed to access the Internet. For example, windows messenger will be used within the real estate company to talk with each other, but all other messengers will be disabled from accessing the external network or Internet to prevent employees from slacking off.

Another form of security that should be performed is a process that divides the network into separate sections. This is also known as "domains of trust." "Domains of trust" is a method used to segment a network into separate parts, based on similar policies and concerns [16]. The server farm and backup servers will be set up within a trusted portion of the network. This will help keep it secure and help keep unwanted users from accessing the information. By placing this information in the private (trusted) zone, it will require: a firewall, user authentication, strong password policies, monitoring, etc. The Internet is placed within the untrusted section of the network, meaning anything can happen in this area so vital information is not available. The last section is called a demilitarized zone or DMZ. This portion of the network will contain and provide the web server and e-commerce server connection to the Internet. This allows users to access certain parts of the network while remaining secure. If the network is attacked through this zone, the attacker only has access to equipment within this DMZ section and not the entire network. As mentioned above, security has to be set in place to protect connections coming into the private network and crossing between each of the specified zones. In order to do this, a firewall will be set in place to filter traffic depending on authorization. Another location that needs to be protected by the firewall and user authentication is the VPN connection. This is a direct connection to the internal network, which needs to remain secure. This will allow the branch office to connect to the main office securely. Not only will the network users need to be authenticated within the trusted network, but also each computer. As mentioned earlier in this document, in order to authenticate each computer, the MAC address filtering should be enabled within the network router to keep out the majority of users. MAC address filtering is design to match an address found on a computer network card with addresses in a list. If the address is found, the computer is allowed to connect to the network; however, if the address is not found, the computer is denied access to the network.

The network administrator will need to monitor the network and the network ports that are left open. These ports need to be constantly monitored to make sure no one breaks into the network and to verify that all needed open ports, work properly.

## 5.4 Acceptable Use of Network Policy

The acceptable use of network policy is designed to inform the network users (both clients and employees) of their responsibilities while connected to the company network. It is to identify areas that can be a risk to the company and what needs to be done in certain situations. By informing the user of the proper usage of company resources, this policy is designed to protect both the company and each user (legally).

The real estate company understands the value of the computer to perform job duties (employees) and to access personal information/research (clients). The company also understands the need for employees to research: real estate tools, templates, auditor information, personal client information, and even background checks. Using a company computer and the computer network (Internet included) is encouraged to help perform the desired job duty and to increase performance. However, by using a company resource, you are subject to the rules defined within this policy. Such resources include: computers, company network, VPN connection, branch network, hardware, software, peripherals, documentation, and even corporate information. The real estate company is given the right to: log or monitor network/Internet usage by the employee/client, monitor data stored on the network/hard drives and has the right to delete any file due to violation without asking for permission, can revoke privileges or delete a user account at any time, decide who can receive company owned equipment such as laptops, and revoke company owned equipment. All data and files stored on the company network remains the property of the real estate company.

Strong passwords should be used to secure each employee's desktop. When the employee navigates away from his or her desk, each employee is responsible to lock the desktop or log off of the network. It is not permitted to log in and then walk away. This is a security risk and a client could easily access private or personal information by using the terminal. Company email accounts will be provided to each employee. The company email address is only meant to be used for business communication purposes and is not designed for personal usage. E-mail chains or forwards, should be sent to a personal email address only. This is to help reduce the amount of viruses or potential phishing attacks on the company network. In addition, e-mail chains or forwards, are a waste of company time and do not benefit the workplace at all. To help secure the information within the company, any equipment connected to or inserted into a company computer/network, needs to be approved ahead of time by the IT administrator. Under no circumstances is an employee or client authorized to engage in any activity that is illegal under local, state, federal or international law while utilizing any company resource.

By using the company network each employee agrees not to perform the following functions. Violate the rights of any person or company, copyrights and patents, laws and regulations, and even intellectual property; use any peer-to-peer software or even download/transmit "pirated" or illegal software; distribute any confidential information pertaining to the real estate company; play online or computer based videogames; or even install any software that has not been pre-approved by the IT department.

The real estate company may make changes to this policy at any time without giving prior notice to the user. Although users will be notified of changes made to the documentation, the user is responsible to check for any changes made to the policy. By violating this policy, the employee's/client's right to network access may be revoked and the employee may/may not be terminated of their employment with the real estate company.

## 5.5 Security Management Policy

The security management policy is designed to find ways to protect company information relating to data, hardware and software. It helps to prevent misuse of the company resources and to protect all information from being lost. To help ensure the security of the network, the administrator needs to monitor the network data to make sure only authorized users are able to view confidential information and the information is not accessible over the Internet. Only authorized users should be allowed to make any changes to the information and the authenticity of the data should be determined. The administrator will need to scan network devices for any errors, such as switches and routers. The connection to/from the VPN connection needs to be tested to make sure the connection is secure. Company systems need to be updated with any new updates, while security devices (firewalls, intrusion detection, and intrusion prevention) are enabled.

Administrators will also need to scan the local network for any problems or vulnerabilities. To avoid some attacks, they can scan the network ports in order to determine how much information is being revealed. This process can be performed by running a program called netstumbler. Another useful tool that can be used within the network to find vulnerabilities is called Microsoft Baseline Security Analyzer (MBSA). This program scans a computer system and displays whether any patches are missing for various products. It will also identify any weak passwords within the computer system.

From a physical security aspect, there is one employee/office throughout the real estate building, which can keep an eye on a client at all times. Although this is a good start, it is not the only security set in place. The server farm itself will be set in the back of the building, within its own room. This room will be locked and the only individuals with access are the CEO and the head IT employee. Although a swipe card can be used to gain access in/out (also to track the employee) of the locked room, it is a little too advanced and over budget for the real estate company at this time. It may be implemented in the future, as long as the company continues to grow. Another practice that will be implemented, especially with dealing with personal client information and private company information, is shredding of all paper. Regular trash can go in the garbage, but any kind of paper will be shredded with a fine crosscut shredder. This is done to protect the client and the corporation.

## 5.6 Incident Handling Policy

The incident handling policy is the last section of a network security policy, in which sets of procedures that should be performed if a great incident occurs are defined. This is usually made up of the legal options and the responsibilities that each person has. One example of a law that helps support the security used within a company is the Canadian Personal Information Protection and Electronic Documents Act (PIPEDA). This attempts to balance the right for a client's private information to be protected and kept confidential, while allowing corporations to obtain/handle the information for business purposes [16]. Almost like the privacy act, this is designed to keep a client's private information confidential. Since the real estate company has a lot of confidential information, this law helps keep the information confidential and protects it from falling into the wrong hands. The company is responsible for this information and can be held liable in case of an accident. Some examples of the private information that the real estate office receives, are: street addresses, phone numbers, social security numbers, pay information, background checks, delinquency information, and etcetera. Because this information can lead to harming the client, theft or even identity theft; there should be standards held within the company. The Canadian Personal Information Protection and Electronic Documents Act law explains this situation and the fact that private information needs to be protected within any organization.

If an incident occurs, the first thing to do is determine what kind of incident it is (attack, misuse, intentional, intrusion, etc.) and the impact it has on the overall security of the network. The incident will need to be documented in order to show what happened, the date it occurred, what it looked like, etc. All physical access to the area will need to be bypassed or shut off, while evidence is being collected. After all of the evidence has been collected, the network will need to be fixed and the administrator will need to determine the underlying reason why/how the incident occurred. If an internal network user is at fault, depending on the severity, they will receive a verbal warning while having their network abilities suspended/revoked. If the user is caught trying to cause harm to the network/stealing, they will be turned over to local authorities. All incidents that are found on the network or through monitoring should be reported to the network administrator immediately.

## 6. Conclusion

Infrastructure security represents one of the most important aspects of an organization, to prevent a breach of confidentiality or the loss/manipulation of data. The best approach when designing a good network security structure is to add components and policies that will help guard the organization, monitor network activity for any unusual activity, test the security vulnerabilities by allowing a trusted source attack the network, and then use the results to enhance the network components/policies. Although testing network vulnerabilities is needed to enhance the security policy, the technique used to test these vulnerabilities can be a contentious topic. By allowing another individual or "trusted" source attack the network, you are providing someone else access to your system and the knowledge of the open vulnerabilities. The knowledge of these vulnerabilities could lead to future attacks against the network or the creation of a backdoor to bypass existing security components. Take the Anonymous hacking group for example. It mainly represents a movement, where each person remains as an individual, but together work for a common cause – security and privacy.

Typically, Anonymous will gather information through any means necessary, in order to provide information to the general mass. Both Anonymous and Lulzsec will break into an organization and steal/post internal information, to prove to the company the lack of security they have implemented and the privatization of information. Each of these groups believes they are helping people by showing them flaws within the system. Yet the release of corporate or government information can potentially cause more harm/destruction to the organization than good. This same idea can be true when testing our own network security. We have to be careful not to leave the security wide open to allow unauthorized access. The less people that know about the security issues the better, in which testing the vulnerabilities should only include trusted sources. In addition, administrators will also need to constantly revise the security components/policies and make constant enhancements to the network. Being proactive when implementing a secure infrastructure will help prevent or minimize internal and external attacks.

## References

[1]     Capella University, (2008). *TS5160 Business Foundations (2nd Custom ed.).* Boston, MA: Pearson Custom Publishing, p. 76.

[2]     Wilson, Z. (2001).        *Hacking:  the  basics.*        Retrieved  on  January  20,  2011  from http://www.sans.org/reading_room/whitepapers/hackers/hacking-basics_955.

[3]     Kryptos      Logic      (2010).      *Konboot.*      Retrieved      on      August      29,      2011      from http://www.kryptoslogic.com/?area=2&item=2.

[4]     Sridhar rao, S. (2011). *Denial of Service attacks and mitigation techniques: real time implementation with detailed      analysis.*      Retrieved      on      November      15,      2011      from http://www.sans.org/reading_room/whitepapers/detection/denial-service-attacks-mitigation-techniques-real-time-implementation-detailed-analysi_33764.

[5]     Manwani, S. (2003).    *ARP cache poisoning and detection.*    Retrieved October 18, 2011 from http://www.cs.sjsu.edu/faculty/stamp/students/Silky_report.pdf.

[6]     McGee, A. R., Vasireddy, S. R., Xie, C., Picklesimer, D. D., Chandrashekha, U., & Richman, S. (2004). A framework for ensuring network security. Lucent Technologies.

[7]     Brotby, W. (2006). *Information security governance: guidance for boards of directors and executive management.* Illinois, IT Governance Institute.

[8]     Jackson, C. (2010). *Network security auditing.* Indianapolis, IN: Cisco Press, pp. 8-9.

[9]     Singh, K. (2001*). IT infrastructure security-step by step.* Retrieved on February 2, 2011 from http://www.sans.org/reading_room/whitepapers/basics/infrastructure-security-step-step_430.

[10]    Bruno, A. & Jordan, S. (2007). *CCDA official exam certification guide [third edition].* Indianapolis, IN: Cisco Press, pp. 42-553.

[11]    SemSim.com (n.d.).    *The cisco three-layeredhierarchial model.*    Retrieved on January 31, 2011 from http://www.mcmcse.com/cisco/guides/hierarchical_model.shtml.

[12]    TechFAQ  (n.d.).        *LAN  switching  and  switch  types.*        Retrieved  January  31,  2011  from http://www.tech-faq.com/lan-switching-and-switch-types.shtml.

[13]    McKelvey, F. (2010). *Ends and ways: the algorithmic politics of network neutrality.* Global Media Journal – Canadian Edition. Volume 3, Issue 1, p. 60.

[14]    Cisco  (n.d.).        *Cisco  catalyst  4500  series  switches.*        Retrieved  February  1,  2011  from http://www.cisco.com/en/US/products/hw/switches/ps4324/index.html.

[15]    Cisco  (n.d.).        *Cisco  2800  series  integrated  services  routers.*        Retrieved  February  8,  2011  from http://www.cisco.com/en/US/products/ps5854/index.html.

[16]    Teare, D. (2008). *Designing for Cisco Internetwork Solutions (DESGN) Second Edition.* Indianapolis, IN: Cisco Press.

[17]    Beamer, Kevin (2007). OSPF. Retrieved on February 20, 2011 from http://searchtelecom.techtarget.com/sDefinition/0,,sid103_gci212728,00.html.

[18]    Stonebumer, G., Goguen, A., and Feringa, A. (July 2002). National Institute of Standards and Technology. *Risk management guide for information technology systems.* Retrieved on February 12, 2011 from http://csrc.nist.gov/publications/nistpubs/800-30/sp800-30.pdf.

[19]    Brandon, J. (2007). *802.11n: fact vs. fiction.* Retrieved on February 21, 2011 from http://archive.laptopmag.com/Features/Common-802-11n-Myths-Debunked.ht.

[20]    CWNP (2003). *Wireless LAN Security Policy Template.* Retrieved February 21, 2011 from http://download.microsoft.com/download/c/6/3/c63d0744-ab83-4aa9-818b-0051b53da77f/ Sample_Security_Plan.doc.

[21]    Rahman, Syed and Donahue, Shannon (2011); "Converging Physical and Information Security Risk Management", Executive Action Series, The Conference Board, Inc. 845 Third Avenue, New York, New York 10022-6679, United States

[22]    Benson, Karen and Rahman, Syed (2011); "Security Risks in Mechanical Engineering Industries", International Journal of Computer Science and Engineering Survey (IJCSES), ISSN: 0976-2760 (Online); 0976-3252 (Print)

[23]    Slaughter, Jason and Rahman, Syed (2011); " Information Security Plan for Flight Simulator Applications"; International Journal of Computer Science & Information Technology (IJCSIT), Vol. 3, No 3, June 2011, ISSN:0975-3826, Vol 3, No 5, Oct 201116

[24]    Jungck, Kathleen and Rahman, Syed (2011); " Cloud Computing Avoids Downfall of Application Service Providers";International Journal of Information Technology Convergence and services (IJITCS), ISSN : 2231 - 153X (Online) ; 2231 – 1939

[25]    Bisong, Anthony and Rahman, Syed (2010); "An Overview of the Security Concerns in Enterprise Cloud Computing "; International journal of Network Security & Its Applications (IJNSA), Vol.3, No.1, January 2011, ISSN: 0975 - 2307

[26]    Halton, Michael and Rahman, Syed (2010); "The Top 10 Best Cloud-Security Practices in Next-Generation Networking"; International Journal of Communication Networks and Distributed Systems (IJCNDS); Special Issue on: "Recent Advances in Next-Generation and  Resource-Constrained Converged Networks"

[27]    Mullikin, Arwen and Rahman, Syed (2010); "The Ethical Dilemma of the USA Government Wiretapping"; International Journal of Managing Information Technology (IJMIT); Vol.2, No.4, November 2010, ISSN : 0975-5586

[28]    Rahman, Syed and Donahue, Shannon (2010); "Convergence of Corporate and Information Security"; International Journal of Computer Science and Information Security, Vol. 7, No. 1, 2010; ISSN 1947-5500

**Authors**

**Kyle Dees** is a Physical Scientist/Analyst for the United States Air Force. He obtained his Bachelor of Science degree, with a focus in physics, from Mount Union College in 2003 and later obtained his Master of Science degree, from Capella University in 2011, focusing on Information Technology with an emphasis in Security. Kyle's research interests include information assurance and security, network administration/development, crowdsourcing, cloud accessibility, and software testing/quality assurance. Kyle has worked in the professional field of information security, software testing and quality assurance for over 10 years.  He is currently A+ certified and Associate Mac Integration 10.6 Apple Certified.

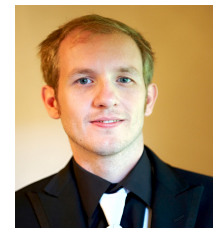

**Dr. Syed (Shawon) M. Rahman** is an Assistant Professor in the Department of Computer Science and Engineering at the University of Hawaii-Hilo and an adjunct faculty of Information Technology, information assurance and security at the Capella University. Dr. Rahman's research interests include software engineering education, data visualization, information assurance and security, web accessibility, and software testing and quality assurance. He has published more than 65 peer-reviewed papers. He is a member of many professional organizations including ACM, ASEE, ASQ, IEEE, and UPE.

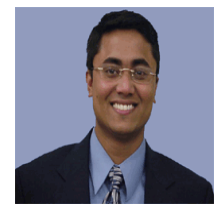